\documentclass [12pt]{article}
\begin{document}
\title{ Tetraquarks  as Diquark-Antidiquark Bound Systems }
\author
{M. Monemzadeh\thanks{monem@kashanu.ac.ir}\\
N. Tazimi \thanks{nt$_{-}$physics@yahoo.com}\\
P. Sadeghi \thanks{ph.parva@yahoo.com}\\
\it\small{{Department of Physics, University of  Kashan, Kashan, Iran.}}}
\date{}
\maketitle
\begin{abstract}
 In this paper, we study four-body systems consisting of diquark-antidiquark, and we analyze diquark-antidiquark in the framework of a two-body (pseudo-point) problem. We solve Lippman-Schwinger equation  numerically for charm diquark-antidiquark systems and find the eigenvalues to calculate the binding energies and masses of heavy tetraquarks with hidden charms. Our results are in good agreement with theoretical and experimental data.
 \end{abstract}

\newpage 
\section{Introduction}
The idea of existence of tetraquark hadrons (two quarks and two antiquarks) was initially raised about twenty years ago by Jaffe. He utilized one of the initial versions of the bag model  to study tetraquark spectroscopy of $q^{2}\bar{q}^{2}$ in which q was a quark lighter than charm quark.  The MIT bag study revealed a dense spectrum of tetraquark states in the light sector \cite{1}. Later, tetraquark systems were examined in potential models and flux tube models \cite{2}. 
The notion of diquark (two-quark system) is of use in describing the hadron structure and particle interactions at high energies. \\

According to the quark model, hadrons are made up of quarks. Mesons consist of a quark and an antiquark in a bound state, such as light scalar mesons and some charmed mesons \cite{3} and baryons are composed of three quarks in a bound state. Their structure was also shown to contain diquarks \cite{4,5,6}. \\

Based on the diquark model, two quarks join and build a color-anti-triplet in a bound state.
As a simplified image, a diquark is viewed as a point particle having the quantum number of two quarks. More generally, a diquark is a system of two quarks considered as a pair. For example, a two-quark correlation in a hadron with at least two quarks will be a diquark system. In its ground state, a diquark has positive parity and may be an axial or a scalar  vector.\\

Gell-Mann \cite{7} originally proposed the existence of  diquarks. Based on the fundamental theory, the concept of diquark was developed, and it was even used to account for some experimental phenomena \cite{8}.  Ref.\cite{9} explored  the exotic state X(3872) via QCD Sum Rules. This study treated the hadronic state as a hidden-charm-tetraquark state with the quantum number $I^{G}(J^{PC}) = 0^{+}(1^{++})$. Chen and Zhu \cite{10} used the same interpolating current to investigate the $1^{+-}$ tetraquark state and found its mass to be (4.02 ± 0.09)GeV.\\

Maiani  et al.\cite  {3} studied the masses of hidden-charm diquark-antidiquark systems taking into account the constituent diquarks' masses and spin-spin interactions, but Ebert et al.\cite{4} employed the relativistic quark model based on the quasi-potential approach in order to find the mass spectra of hidden-charm tetraquark systems. Unlike Maiani  et al., they ignored the spin-spin interactions inside diquark and inside anti-diquark. We, however, considered tetraquarks as two-body systems, and spin-spin interactions were ignored. We used the diquark-antidiquark picture to reduce a complicated four-body problem to two simpler two-body problems. The paper is organized as follows:\\
In the first part, the  bound states of four-quark systems are investigated in the framework of a pseudo-point two-body system. We explain Gauss-Legendre method in the second part. In the third part, we calculate the binding energy of heavy tetraquarks with hidden charms, and the last part is devoted to conclusions.\\

\section{Tetraquarks represented through two-body problems}
 Among the tetraquark states, those consisting of diquark-antidiquark are of interest to this study. To describe tetraquarks, we have taken diquark-antidiquark as if the pseudo-point diquark was a two-body system. Such interpretation helps reduce a complex relativistic problem to a simple two-body problem (Fig.1).\\
 
\begin{figure}[h]
\includegraphics{tetra-pic.jpg}
\vspace{5cm}
 \caption{Tetraquark systems }
\label{1}
\end{figure}
\vspace{0.5cm} 
 Due to the wide variety of heavy diquarks, we have narrowed the study down to hidden-charm diquarks. 
 These particles consist of a charm quark and a light quark ($u,d,s$).  In order to use unrelativistic Schrodinger equation with tetraquark systems, we take heavy diquark and solve homogeneous Lippman-Schwinger equation numerically \cite{11,12} (which is the integral form of Schrodinger equation) for two-body systems composed of diquark-antidiquark .\\

Schrodinger equation for a two-body bound state with the potential $V$ (which is assumed to be energy-independent) runs as the following integral equation \cite{11}:
\begin{equation} 
\mid\psi_{b}>=G_{0}V\mid\psi_{b}> 
\label{1}
\end {equation}
$G_{0} $ is the propagator of a free particle. In configuration space, Eq.(\ref{1}) turns out as:
\begin{equation} 
\psi_{b}(r)=-m\frac{1}{4\pi}{\int_{0}^{\infty}}{ dr^\prime} {r^\prime}^{2}{\int_{-1}^{1}} dx^\prime {\int_0}^{2\pi} d\phi ^\prime \frac{exp(-\sqrt{m\vert E_{b}\vert} \vert{r-r^\prime}\vert)}{\vert{r-r^\prime}\vert} V(r^\prime){\psi_{b}(r^\prime)} 
\label{2}
\end {equation}
where $ E_{b} $ stands for the binding energy of the two-body bound system (diquark+antidiquark).  The wave function will be:
\begin{equation}
{\psi_{b}(r)}=\int_{0}^ {\infty}dr^{'} \int_{-1}^ {1} dx^{'} M(r,r^{'}, x^{'}) {\psi_{b}(r^{'})} 
\label{3}
\end{equation}
where:
\begin{equation} 
M(r,r^{'},x^{'})=-2\pi m \frac{1}{4\pi} \frac{exp((-\sqrt{m\vert E_{b}\vert) }\sqrt{r^{2}+\acute{r}^{2}-2r \acute{r} \acute{x}})}{\sqrt{r^{2}+\acute{r}^{2}-2r\acute{r} \acute{x}}}{} \\ 
{ V(\acute{r}^{2})} 
\label{4}
\end{equation}
Eq.(\ref{4}) is of the following eigenvalue form:
\begin{equation}
K(E_{b})\vert\psi_{b}>=\lambda(E_{b})\vert\psi_{b}>
\label{5}
\end{equation}
To determine the binding energy, we  diagonalize the kernel in our calculations to obtain an eigenvalue   $ \lambda=1  $, which is indicative of a physical system. The masses of the constituent quarks are as follows:
\begin{equation}
m_{s}=0.5 GeV\qquad m_{u}=m_{d}=0.33 GeV\qquad m_{c}=1.55 GeV
\end{equation}
The other required data include an arbitrary potential,  $r$-cutoff,  and reduced mass of the diquark-antidiquark. $R$-cutoff is supposed to be the point at which the potential tends to zero.\\

\section{Gauss-Legendre  method}
 The eigenvalue equation (5) is solved through iteration method (direct method) \cite{13}. To discretize the integrals, Gauss-Legendre method \cite{14} is employed. Gauss lattice points for $r,  r \prime, x\prime$ are supposed to be $100$. In Gauss-Legendre method, each integral of [-1,+1] interval is treated as:
\begin{equation}
\int _{-1}^{+1} f(x) dx=\sum^{n}_{i=1} w_{i} f(x_{i})
\label{5}
\end{equation}
where  $x_{i}$ denotes the roots of the type-one order-N Legendre function, and $w_{i}$ are the functions of point weight. The following variable change is used to transfer the  integration interval of $r \prime$ from $[0, r_{max}]$ to $[-1,+1]$. If the integrals are discretized, then:
\begin{equation}
r =r_{max} \frac{1+x}{2} 
\label{5}
\end{equation}
\begin{equation}
{\psi_{b}(r)}=- 2\pi m\sqrt{\pi /2} \sum_{j=1}^{N_r^\prime} \sum_{i=1}^{N_r^\prime}{W_{r_{i}}^\prime} {W_{x_{j}}^\prime} {r^\prime_{i}}^{2}\frac{exp(-\sqrt{m\vert E_{b}\vert} \rho(r,r^\prime_{i},x^\prime_{j} ))}{\rho(r,r^\prime_{i},x^\prime_{j} )} V(r_{i}^\prime){\psi_{b}(r^\prime_{i})}
\label{6}
\end{equation}
Eq.(\ref{6}) could be reformulated as:
\begin{equation}
{\psi_{b}(r)}=\sum_{i=1}^{N_{r^{'}}} N(r,r^{'}_{i}) {\psi_{b}(r^{'}_{i})}
\label{7}
\end{equation}
where:
\begin{equation}
N(r,r^\prime_{i})=-2\pi m\sqrt{\pi /2} \sum_{j=1}^{N_r^\prime} {W_{r_{i}}^\prime} {W_{x_{j}}^\prime} {r^\prime_{i}}^{2}\frac{exp(-\sqrt{m\vert E_{b}\vert} \rho(r,r^\prime_{i},x^\prime_{j} ))}{\rho(r,r^\prime_{i},x^\prime_{j} )} V(r_{i}^\prime)
\label{8}
\end{equation}
Matrix $N$ is diagonized to find $\lambda=1$ in the eigenvalue spectrum. The energy corresponding to $\lambda=1$ will be the system's binding energy.\\

\section{ Binding energy and masses  of heavy tetraquarks with hidden charms }
In this part, we use a potential presented in \cite{15}. In this potential, diquark interaction results from gluon field (spin-spin interaction in the potential is ignored). The potential is of this form:
\begin{equation}
V(r)=V_{coul}(r)+V_{conf}(r)\\
\end{equation}
\begin{equation}
V_{coul}(r)=-\frac{4}{3}\alpha_{s}\frac{F_{1}(r)F_{2}(r)}{r}\\
\end{equation}
\begin{equation}
V_{conf}(r)=Ar+B
\end{equation}\\
where $F(r)$ is the form factor, which enters the vertex of the diquark-gluon interaction.\\
\begin{equation}
F(r)=1-e^{\xi r-\zeta r^{2}}
\end{equation}
$A=0.18 GeV^{2}, B=-0.3 GeV$, and $\alpha_{s}$ is the strong coupling constant. The masses and parameters $\zeta$ and $\xi$ for [c, q] and $\lbrace c, q\rbrace$ are given in Table 1. S and A denote scalar and axial vector diquarks of antisymmetric $[c, q]$ and symmetric $\lbrace c,q\rbrace$ in flavor, respectively.\\

The diquark is not indeed color-singlet, nor is it a real physical state. It generally exists in baryons. Thus, the estimated mass of a free diquark may differ from its actual mass in baryons. It is because of QCD interactions of the extra quark with the diquark \cite{16} .\\

 Kleiv et al. \cite{17} discussed the uncertainty of mass prediction  of charm-light diquarks as resulting from the uncertainties in QCD parameters. They presented the lower bound of $[ c,q]$ mass in two states $0^{+}$ and $1^{+}$ as $1.86\pm0.05$ and $1.87\pm0.10$, respectively, while they presented the upper bound of the mass as 2.02 and 2.07 for $0^{+}$ and $1^{+}$ states, respectively.\\

\begin{small}
\begin{center}
Table 1: Masses  and form factor parameters of heavy-light diquarks \cite{18}.
\end{center}
\begin{center}
\begin{tabular}{l c c c c c p{5cm}l}
\hline
 Quark content & Diquark type & $M$(MeV) & $\xi$ (GeV) &$ \zeta$ ($GeV^{2}$)\\
\hline 
 [c,q] &S  & 1973&2.55 &0.63 \\ 
\hline
$\lbrace c,q\rbrace$ &A  & 2036&2.51 &1.05  \\ 
\hline
 $[c,s]$ &S  & 2091&2.15 & 1.05 \\ 
\hline
$\lbrace c,s\rbrace$ & A &2158 &2.12 & 0.99\\ 
\hline 
\end{tabular}  
\end{center}
\end{small}
\vspace{1.5cm}
A number of theoretical approaches have been proposed to study heavy diquark masses. Examples include Bethe-Salpeter equation \cite{18}, constituent diquark model \cite{19}, and relativistic quark model based on a quasipotential approach in QCD \cite{20}. It is worth mentioning that diquark states in the nuclear matter are of masses larger than those states in the vacuum. Ref. \cite{21} presents these mass uncertainties for light-flavor diquark states.\\

 Inserting the data given in Table 1 and diagonalizing the kernel and finding $ \lambda=1 $ eigenvalue, we can calculate the binding energy. Using these results and the relation of mass and binding energy in Eq.(16), we obtain tetraquark system masses.
\begin{equation}
M=m_{1}+m_{2}+\frac{E_{b}}{c^{2}}
\end{equation}
where $ m_{1} $,$ m_{2} $ are diquark and antidiquark masses, respectively. In Table 2 and Table 3,  binding energies and masses  for charm tetraquarks are presented. The masses obtained via this method turn out to be in such good agreement with the experimental and theoretical masses  that the observed errors are insignificant. In Fig.2 and Fig.3 we have compared our results with Maiani et al.  and Ebert et al.\\

\begin{small}
\begin{center}
Table 2: $cq\bar{c}\bar{q}$ Masses and binding energies obtained via our method compared with theoretical predictions and possible experimental candidates in different $ J^{PC} $ states
\end{center}
\end{small}
\begin{tabular}{l c c c c c c c p{5cm}l}
\hline Tetraquark & $ J^{PC} $ & The calculated   & The calculated & Mass in   & Mass in& Exp\\ 
& & $E_{b} (MeV)$&  mass  (MeV) & ref \cite{3,25} & ref \cite{15} \\
 \hline
$S\bar{S} $  & $0^{++}$ & -242.86  &3703.14  &3723  &3812 &  \\ 
$\frac{(S\bar{A}+ A\bar{S})}{\sqrt{2}}$ &$1^{++}$  & -147.32 & 3861.68 & 3872   &3871&X$\lbrace^{3871.4 \cite{22}}_{3875.2 \cite{22}}$ \\  
$\frac{(S\bar{A}+ A\bar{S})}{\sqrt{2}}$ &$1^{+-}$  &-255.2  &3744.84  &  3754&3871 &  \\ 
$A\bar{A} $&$0^{++}$  &-232.92  &3839.08  &3832  &3852 &  \\ 
$A\bar{A} $&$1^{+-}$ &-185.8 &3886.2  &3882  &3890 &  \\ 
 $A\bar{A}$ &$2^{++}$  &-135.25 &3946.75 &3952  &3968 & Y$\lbrace^{3943 \cite{23}}_{3914.3 \cite{24}}$\\ 
\hline 
\end{tabular}\\
\\
 
\vspace*{1.1cm} 
\begin{small}
\begin{center}
Table 2: $cs\bar{c}\bar{s}$ Masses and binding energies  predicted via our method compared with theoretical results in different $ J^{PC} $ states\\
\vspace*{0.75cm}
\begin{tabular}{l c c c  c c l}
\hline Tetraquark & $ J^{PC} $ & The calculated   & The calculated & Mass in  \\ 
& & $E_{b} (MeV)$&  mass  (MeV) & ref \cite{15} \\
 \hline
$S\bar{S} $  & $0^{++}$ & -128.1  &4053.9  &4051    \\ 
$({S\bar{A}+ A\bar{S}})/ {\sqrt{2}}$ &$1^{++}$  & -127.29 & 4111.71 & 4113 \\  
$({S\bar{A}-A\bar{S}})/{\sqrt{2}} $ &$1^{+-}$  &-150.34  &4098.66  &  4113  \\ 
$A\bar{A} $&$0^{++}$  &-216.38  &4099.62  &4110   \\ 
$A\bar{A} $&$1^{+-}$ &-180.47 &4135.53  &4143    \\ 
 $A\bar{A}$ &$2^{++}$  &-101.97 &4214.03 &4209  \\ 
\hline 
\end{tabular}\\

\end{center}
\end{small}
\vspace*{1.5215cm}
It must be notified that we did not introduce any new or different parameters in potential for calculating the mass spectrum of heavy tetraquarks but employed the values already presented in \cite{15, 18} and obtained the tetraquark mass spectrum by the two-body system's binding energy. A good agreement was observed between our results and experimental data and other references \cite{15,3,25}. Therefore, we can be confident that the potential coefficients chosen from   ref.[15]  in our calculations to solve Lippman-Schwinger equation were appropriate. Furthermore, we can conclude that our proposed binding energy was appropriate because it led to results nearly the same as those obtained by other scholars in previous studies.
 In Fig.2 and Fig.3, we have compared our results with other theoretical results  for these systems.\\

\vspace{2cm}
\begin{figure}[h]
\includegraphics{20.png}
\vspace{4cm}

\begin{small}
\caption{Mass spectrum of $cq\bar{c}\bar{q}$. a:our work, b:Ebert  et al.'s work, c: Maiani et al.'s work  }
\label{2}
\end{small}
\end{figure}
\vspace{0.6cm}
 In the recent work of Maiani  et al. (‘type-II’ diquark model), diquarks are more resembling compact bosonic building blocks \cite{26}. They have  considered only diquark-antidiquark spin interactions and thus they have neglected spin-spin interactions between different diquarks. This means these results are an approximation of ref\cite{3}. 
  Ignoring the spin interactions within diquarks and within antidiquarks  and using different K-coefficients have contributed to the differences between the numerical value of the masses from experimental results and from their previous results \cite{3}.\\
  
  The potential used in the paper is spin-independent. The mass differences observed in different $J^{PC}$ states result from the different masses and form factor parameters of heavy-light diquarks that we used for these states (Table 1).
Thus, spin indirectly affects the tetraquark systems' binding energies and masses (Table 2).
Therefore, the mass differences observed between our results and those of ref.[15] derive from the fact that we ignored spin interaction in the potential. In our future works, we will add spin to the potential.
\newpage
\vspace{8.5cm}
\begin{figure}[h]
\includegraphics{21.png}
\vspace{11.95cm}
\begin{small}
\caption{Mass spectrum of $cs\bar{c}\bar{s}$. a: our work, b: Ebert et al.'s work }
\label{2}
\end{small}
\end{figure}
\vspace{0.5cm}
\vspace{0.04cm}
\section{Conclusion}
In this paper, we made use of the potential coefficients proposed by Ebert et al. and solved Lippman-Schwinger equation for heavy tetraquark systems. We managed to obtain the binding energy and used it to calculate the masses of heavy charm tetraquarks. The tetraquark is considered as the bound state of a heavy-light diquark and antidiquark. We used the diquark-antidiquark picture to reduce a complicated four-body problem to two simpler two-body problems. 
 Our masses are very close to experimental and theoretical masses. Thus our method is appropriate for investigating heavy tetraquarks. Our results are in good agreement with the results derived from complicated relativistic methods and can be a good replacement for them.\\
This method could equally be used for bottom tetraquarks   and the tetraquarks composed of two heavy quarks and two heavy antiquarks.\\

\section{Acknowledgments }
We are pleased to thank the University of Kashan for Grant No. 65500.4.

\end{document}